# A carbon-nanofiber glass composite with high electrical conductivity

Guangming Tao[1,2†], Shi Chen[1†], Sudeep J. Pandey[1], Felix A. Tan[1], Heike Ebendorff-Heidepriem[3], Michael Molinari[4], Ayman F. Abouraddy[1,5*], and Romain M. Gaume[1,5,6*]

[1]CREOL, The College of Optics & Photonics, University of Central Florida, Orlando, Florida 32816, USA

[2]Wuhan National Laboratory for Optoelectronics and School of Optical and Electronic Information, Huazhong University of Science and Technology, Wuhan 430074, China

[3]Institute of Photonics and Advanced Sensing, School of Chemistry and Physics, ARC Centre of Excellence for Nanoscale BioPhotonics, The University of Adelaide, Adelaide SA 5005, Australia

[4]CBMN UMR CNRS 5248, University of Bordeaux, Bordeaux INP, 33600 Pessac, France

[5]Department of Materials Science and Engineering, University of Central Florida, Orlando, Florida 32816, USA

[6]NanoScience Technology Center, University of Central Florida, Orlando, Florida 32816, USA

[†]Equal contribution

[*]Corresponding authors: raddy@creol.ucf.edu, gaume@ucf.edu

## Abstract

The use of oxide glasses is pervasive throughout everyday amenities and commodities. Such glasses are typically electrical insulators, and endowing them with electrical conductivity – without changing their salutary mechanical properties, weight, or thermoformability – enables new applications in multifunctional utensils, smart windows, and automotive parts. Previous strategies to impart electrical conductivity include modifying the glass composition or forming a solid-in-solid composite of the glass and a conductive phase. Here we demonstrate – using the latter strategy – the highest reported room-temperature electrical conductivity in a bulk oxide glass (~ 1800 S/m) corresponding to the theoretical limit for the loading fraction of the conductive phase. This is achieved through glass-sintering of a mixture of carbon nanofibers and oxide flint (F2) or soda lime glasses, with the bulk conductivity further enhanced by a polyethylene-block-poly(ethylene glycol) additive. A theoretical model provides predictions that are in excellent agreement with the dependence of conductivity of these composites on the carbon-loading fraction. Moreover, nanoscale electrical characterization of the composite samples provides evidence for the existence of a connected network of carbon nanofibers throughout the bulk. Our results establish a potentially low-cost approach for producing large volumes of highly conductive glass independently of the glass composition.

# 1 | Introduction

Glasses offer a unique set of mechanical and optical properties that underlie their wide-spread utilization in everyday commodities[1-4]. The strength of glasses – in conjunction with thermoforming processing, low density, and low cost – facilitates their utilization in large-area windows, enables thermal drawing into extended fibers for optical telecommunications, and drives a plethora of applications from cooking utensils to screens for electronic devices and the encapsulation of solar panels. Glasses are typically electrical insulators. Notable exceptions include chalcogenide glasses[5] – that are in fact amorphous semiconductors[6] – and recently developed metallic glasses[7, 8]. The most widely used oxide glasses, however, feature extremely low conductivities ~ $10^{-15}$ – $10^{-10}$ S/m. Developing an electrically conductive oxide glass would have profound implications for the automotive industry[9], thermal cloaking[10], and novel multifunctional multimaterial fibers[11-14]. Metals are not a viable substitute for a conductive glass in these application areas because of their higher density, thermal conductivity, and thermal expansion coefficients, and their crystallinity implies a sharp transition in viscosity above the melting temperature, in contrast to glasses that can be processed thermally in the viscous state. Moreover, the low operating temperatures of conductive polymers preclude their utility in many applications.

A significant quest for increasing the electrical conductivity of oxide glasses without diminishing their other salutary characteristics has therefore been pursued. To date, two strategies have been explored to endow oxide glasses with electrical conductivity. One strategy relies on modifying the chemical composition of the glass, usually through the inclusion of metallic elements in the glass matrix, with reported ionic conductivities[15, 16] ~ $10^{-3}$ – $10^{-1}$ S/m. A second strategy makes use of a solid-in-solid composite in which the glass matrix is loaded with a conductive phase – typically one of the various forms of carbon, such as carbon nanotubes (CNTs)[9,17,18], graphitic nanoparticles[19], carbon nanofibers (CNFs)[20], or graphene[21].

Here, we exploit the ceramic-inspired process of glass sintering to produce bulk oxide-glass composites that retain the mechanical properties of glass while featuring a high room-temperature electrical conductivity by loading CNFs into the glass matrix. One-dimensional CNFs favor establishing a connected network compared to nanoparticles, they do not hinder the glass flow as do two-dimensional graphene or graphite, and they have significantly lower costs than CNTs. Using loading concentrations ranging from 1% to 46% by volume of CNFs that are efficiently dispersed in the glass matrix via a non-ionic surfactant, we obtain electrical conductivities in bulk samples as high as ~ 1800 S/m at room temperature, reaching the theoretical limit set by the carbon-loading fraction. Our highest achieved value exceeds the highest reported conductivities for a carbon-loaded oxide glass by more than a factor of ~ 40.[22] Measurements of the dependence of the measured conductivity on the carbon-loading volume-fraction reveals two regimes: tunneling-based conductivity at low carbon-loading followed by percolation above a carbon-loading threshold of ~ 2.3 vol%. A theoretical model based on the global tunneling network model provides predictions that are in excellent agreement with the measurements. Furthermore, the measured temperature-dependence of the conductivity reveals a clear signal for thermally-assisted variable range hopping for low carbon-loading and direct conduction with increased carbon-loading. The existence of a connected network of CNFs throughout the bulk composite is confirmed through nanoscale electrical measurements performed via Kelvin probe force microscopy. Finally, the universality of this approach is demonstrated by utilizing the same approach to create composites based on two common glasses: flint (F2) and soda lime glasses, with the type of glass determining the background conductivity and the percolation threshold. These results provide a general framework for large-scale, low-cost production of novel electrically conductive oxide glass composites.

## 2 | Materials selection for producing glass-CNF composites

The first question to tackle is that of the dimensionality of the conductive carbon phase to be selected. Nanoparticles (zero-dimensional) are less likely to provide a connected network except at a high loading fraction, whereas two-dimensional phases such as graphite or graphene can hinder the glass flow. We have thus settled on utilizing one-dimensional carbon phases, such as CNTs or CNFs that have been utilized as filler materials in a wide range of composites to provide electrical conductivity and mechanical strengthening. Such composites typically perform far below the theoretical expectations because of the attractive van der Waals forces that promote highly entangled agglomerations of CNTs and CNFs, thereby preventing their uniform dispersion. According to the Derjaguin–Landau–Verwey–Overbeek (DLVO) theory, particles can be dispersed when electrostatic repulsions – induced by similarly charged electrical layers surrounding the particles – produce an energy barrier that overcomes the attractive van der Waals potential and opposes aggregation[23,24]. If the magnitude of this energy barrier exceeds the kinetic energy of the particles, the suspension is stable. In light of these requirements, low-cost CNFs offer several salutary features compared with CNTs: CNFs have larger dimension and numerous defect planes along the surface, thus presenting a lower attractive surface energy to overcome[25] and facilitating surface-modification via surfactant adsorption. These factors contribute to a potentially more efficient dispersal of single CNFs in a solution compared to CNTs[26]. We make use of a non-ionic surfactant – an ethylene oxide derivative, Polyethylene-block-poly(ethylene glycol), or PEPEG – to prevent charge attraction[23]. The Poly-(ethylene oxide)-based head groups tend to be located in the hydrophilic phase (ethanol and glass particles), affording a similar nature to the CNFs. The solvent and glass particles therefore lower the attractive potential between the CNFs, while the hydrocarbon tails react physically with the CNF surface. In this scenario, longer hydrocarbon tails are preferred and block copolymer surfactants can create high steric stabilization and prevent the CNFs from approaching each other[27,28].

## 3 | Preparation and characterization of the glass-CNF composites

Traditional ceramic processing (powder-mixing followed by milling and hot-pressing) was utilized to produce glass-CNF composites[29] (Fig. 1a). In one class of samples, bulk F2 glass (CAS-number 65997-17-3, Schott) was crushed and ground into a fine powder using an alumina mortar and pestle. The glass powder was thoroughly mixed for 24 hours with varied CNF loading fractions ($\phi \approx$ 1, 2, 3.7, 7, 10, 16, 36, and 46 vol%) in a planetary ball-mill using zirconia balls for grinding and ethanol as suspension medium. We added 5 wt% of PEPEG as a surfactant to promote the dispersion of the CNFs in the slurry. After drying at 60 °C in a vacuum oven, the well-dispersed glass-CNF powder mixture was loaded into a graphite die and sintered into 25-mm-diameter, 4-mm-thick pellets using hot-press sintering at a temperature of 450 °C and a uniaxial pressure of 100 MPa under a low vacuum (150 mtorr) for 2 hours (Fig. 1b). The densification of composites by hot-pressing is dominated by viscous flow which causes the softened glass phase to flow and wrap the embedded CNFs. The microstructures of the polished surface of glass-CNF composite samples with low carbon-loading fraction by volume $\phi$ can be fully densified through this method (Fig. 1b). The microstructure of the polished surface of a glass-CNF composite at $\phi = 10$ vol% showed no porosity in the sample and demonstrated the full densification of the composite after hot-press sintering (Fig. 1c). In the Supplemental Materials we present the corresponding results that make use of soda lime glass in lieu of F2 glass in the glass-CNF composite.

Varying the CNF loading factor $\phi$ in the powder mixture (Fig. 2a) enables controlling the conductivity of the sintered composite. Fracture-surface microstructures of composites with different $\phi$ (Fig. 2b-i, corresponding to $\phi = 1$ vol% to 46 vol%) are shown in Fig. 2 to illustrate the distribution of CNFs in the glass matrix. Scanning Electron Microscopy (SEM, Zeiss Ultra-55) was used to characterize the sintered composite microstructures, and further analysis of elemental carbon distribution in these samples (on polished surface for less noise) was performed using Energy Dispersive Spectroscopy (EDS; Noran System 7 equipped with a silicon drift detector), as shown in the insets of Fig. 2. The microstructures obtained on the fracture surfaces with increasing $\phi$ reveal that the distribution of CNFs in the glass matrix is homogeneous, as further confirmed by EDS carbon mapping. As expected in the semi-quantitative elemental mapping, the carbon element appears denser in the composite as $\phi$ increases. From these images, the diameter of the CNFs can be estimated to be ~ 130 nm, significantly larger than that of CNTs. The actual length of the CNF cannot be determined from these images as most of them are still embedded in the glass matrix or may have been damaged by the fracture process. The black worm-like voids (for example seen in the $\phi = 3.7$ vol% sample, Fig. 2c) are hypothesized to be gaps left by the pulling of CNFs during fracture. Finally, for $\phi > 36$ vol%, the scarcity of the glass phase prevents the formation of compact composites and some porosity remains.

We measure the electrical conductivity $\sigma$ of the bulk glass utilizing samples in the form of parallelepiped shaped bars (Fig. 3a, inset) via the four-point probe technique (Supplemental Materials). Measurement results of room-temperature $\sigma$ of F2-CNF glass composites are presented in Fig. 3, where different stages are clearly visible in the evolution of $\sigma$ as a function of the $\phi$. See Supplemental Materials for the corresponding measurements on the soda-lime-glass-based composites. The high electrical conductivity of the composite achieved here (~ 1800 S/m at high $\phi$) results in part from the homogeneous distribution of the CNFs in the glass matrix, thereby creating a well-connected conductive network (further confirmed below via nanoscale conductivity characterization). While high $\phi$ result in much higher conductivity, the lower density and fragility of the composites may limit their applicability. These considerations come into play in the selection of a proper value of $\phi$ to ensure a set electrical conductivity value with adequate manufacturability.

## 4 | Modeling the conductivity of glass-CNF composites

In this Section we provide a theoretical model that accounts for the measured behavior of the electrical conductivity presented in Fig. 3. At low values of $\phi$, limited conductivity is obtained through field-induced tunneling (FIT) between distant CNFs[30]. A simple approach assumes that two adjacent fibers form a connected electrical path if their separation is lower than a tunneling length $\xi$ (typically on the order of a few nanometers[31]). Within this framework, an increase in $\phi$ gives rise to a percolation transition between macroscopic insulating and conducting states[32, 33] at a critical loading fraction $\phi_c^p$. With further increase in $\phi$, the effective size of interconnected fiber bundles diverges rapidly to form a continuous network that spans the entire sample, resulting in an increase of conductivity by several orders of magnitude and ultimately reaching that of a compact CNF pellet. In the vicinity of the percolation transition when $\phi > \phi_c^p$, renormalization group theory establishes that $\sigma$ follows the scaling law[34]:

$$\sigma = \sigma_0 (\phi - \phi_c^p)^t, \quad (1)$$

where $\sigma_0$ is a constant and the critical exponent $t$ (typically 1.6 – 2.0 in three-dimensional systems[34]) is thought to be universal, independently of the composite characteristics. Using this description, the room-temperature $\sigma$ data of our composites in Fig. 3 yields $t = 1.7$, an exponent well within the expected universal range, $\phi_c^p = 2$ vol%, and $\sigma_0 = 10^4$ S/m.

The value of $\phi_c^p$ is essentially determined by the relative arrangement and the dimensions of the fibers (effective length $L$ and diameter $D$). In particular, a uniform random dispersion combined with a high fiber aspect ratio $L/D$ favor a percolation onset at low $\phi$. This trend is well captured by Kusy-type relationships[35] for a random distribution of conductive sticks[36, 37], $\frac{L}{D} \phi_c^p \sim \frac{3}{2}$. With $\phi_c^p = 2$ vol%, this relationship predicts a fiber aspect ratio of $L/D \sim 79$, a value consistent with our SEM observations ($D \sim 130$ nm, $L \sim 15$ μm, $L/D = 115$). By extrapolating Eq. 1 to $\phi = 100$ vol%, one estimates the conductivity of a compact pellet of CNF to be $\sigma(\text{CNF}) = 9.7 \times 10^3$ S/m. This value is twice that previously reported for CNT-pellet conductivity ($5 \times 10^3$ S/m).[38-43]

Despite the elegant form of the previous description and the agreement between the data and the theoretical predictions for $\phi > \phi_c^p$, the data nevertheless diverges at low $\phi$ from the scale-invariance inherent to this model. This is readily explained via the global tunneling network (GTN) model, which takes into consideration that the short-range tunneling conduction does not imply a sharp cut-off in inter-particle connectivity. The introduction of a tunneling characteristic length $\xi$ breaks the scale-invariance of conductivity with respect to the inter-particle distance $\delta_c$ (and thus with respect to $\phi$). This departure from scale-invariance dominates at low $\phi$ (whereupon $\delta_c > \xi$). By accounting for the particle shapes and dimensions explicitly, this approach reproduces the overall trend of Eq. 1 for a wide range of composites[40]. Specifically, at low $\phi$ (up to a few vol%), we have $\delta_c \sim \frac{0.6}{(L+D)} \frac{D^2}{\phi}$, which establishes the relationship

$$\sigma \sim \sigma_i + \sigma_0 \exp\left(-\frac{2\delta_c}{\xi}\right). \quad (2)$$

Here, $\sigma_i$ is the bulk conductivity of the insulating phase and $\sigma_0$ is a constant. In the absence of inter-particle conductance cutoff, the GTN model predicts that the insulator-conductor transition results from the tunneling conductivity on the loose fiber network overcoming that of the bulk conductivity of the insulating phase at a filling content $\phi_c^{\text{GTN}}$:

$$\phi_c^{\text{GTN}} = \frac{0.6 D^2}{\xi(L+D)} \frac{1}{\ln(\sigma_0/\sigma_i)}. \quad (3)$$

Fitting our data to this model yields $\xi = 5.9$ nm, $\phi_c^{\text{GTN}} = 1.3$ vol%, $\sigma_0 = 924$ S/m and $\sigma_i = 10^{-3}$ S/m (Fig. 3). Note that reduction of $Pb^{2+}$ in the F2 glass to metallic lead can potentially increase $\sigma_i$.[44] Worth referring to this work by Blodgett in 1951. The tunneling length matches the value of $\xi = 5.9$ nm obtained for CNF in Ref. [45], while the crossover value $\phi_c^{\text{GTN}}$ is consistent with that obtained in the percolation model $\phi_c^p$.

Compared to our measured value of the bulk conductivity of a pure F2 glass sintered sample $\sigma(F2) \sim 10^{-8}$ S/m, the large value of $\sigma_i$ obtained through this fit most certainly arises from partial carbo-reduction of the glass during the sintering process. In fact, it was observed that, when pure F2 glass is processed in the presence of dispersant alone, the bulk conductivity of the sintered sample increases dramatically to $\sigma(F2 + PEPEG) = 2.9 \times 10^{-4}$ S/m, a value comparable to $\sigma_i$.

## 5 | Nanoscale characterization

The theoretical model described in the previous Section yields predictions that are in excellent agreement with the observed $\phi$-dependent room-temperature conductivity of the glass-CNF-composite. Most notably, this model posits the existence of a connected network of CNFs above the percolation threshold. We now proceed to describe nanoscale measurements of our samples that identify the underlying electrical conduction mechanisms and ultimately lends support to the underlying hypothesis. Towards this end, we rely on Kelvin Probe Force Microscopy (KPFM). A first class of measurements were performed by electrically polarizing the sample with two close contacts provided by external tips that can be displaced on the surface to examine the electrical potential change within an area of interest (Fig. 4a). The distances between the contacts in this configuration are ~ 150 – 300 µm (smaller distances provoked short-circuit damage) and the potential difference was ~ 2 V. To obtain the spatial distribution of the electric potential across the sample surface, an AFM tip scans the surface at a given frequency that is kept constant thanks to the application of a supplementary voltage corresponding to the surface potential. The experiments were repeated on 5 different areas of each sample to confirm the reproducibility of the measurements. For a reference CNF-free sample (F2 glass only), the potential drop is homogeneous on a small area and drops uniformly across the scanned area except for abrupt features that are likely due to the surface topography (Fig. 4b). When the same experiment was performed on a F2-CNF composite, the recorded potential distribution was markedly distinct (Fig. 4c). And the results were same from different spots as regular representation of the whole sample. Indeed, we observe the appearance of clear spots of higher potential against a uniformly dropping background. These spots are randomly distributed on the surface in a manner reminiscent of those in the SEMs of surface fractures shown in Fig. 2, and are not related to topographic structures or sample roughness. This different nanoscale conductive behavior reveals the existence of areas of high conductivity compared to the reference sample that we hypothesize is the result of the CNFs emerging at the sample surface.

To corroborate this hypothesis and – furthermore – to test the hypothesis of a conductive network throughout the bulk sample, we carry out a second class of nanoscale electrical-conductivity characterization. Here we perform conductive-AFM measurements in which the AFM tip is in contact with the sample surface while a potential is applied between the AFM tip and the chuck in contact with the sample *bottom* surface so that I-V curves of the bulk sample can be acquired locally (Fig. 4d). The measured I-V curves are expected to be related to the nature of contact between the tip and the sample surface. If the contact is electrically weak (or the tip is in contact with dielectric part of the sample), then we anticipate a Schottky contact and thus a non-ohmic behavior for the I-V curve. On the other hand, contact with a conductive part of the sample (a CNF) will likely display an ohmic I-V curve if a conductive network is spread throughout the bulk. By performing these measurements on CNF-free glass samples and F2-CNF composites, we observe distinct behavior. In the reference CNF-free sample, non-ohmic contact was observed in all the examined areas (Fig. 4e). Measurements on the F2-CNF composite reveal clear ohmic contacts at the conductive spots identified in Fig. 4c and non-ohmic contacts from the rest of the sample (Fig. 4f). Moreover, the measured current in the case of ohmic-contact is ~ 6 orders-of-magnitude larger than the current in the non-ohmic case. The fact that we obtain ohmic behavior by scanning the whole depth of the sample indicates the existence of conductive paths inside the bulk composite glass and not just at the surface.

# 6 | Temperature-dependence of the composite conductivity

We have also investigated the temperature dependence of the electrical conductivity for the F2-CNF composites with different carbon-loading fractions $\phi$. The measured temperature dependence of $\sigma$ in the F2-CNF composites is shown in Fig. 5a. Measurements were performed over the range of temperature values 293 – 673 K, below the glass transition temperature of F2 glass ($T_g = 707$ K). In this range, measurements reveal that the conductivity increases with increasing temperature ($d\sigma/dT > 0$), a behavior indicative of carrier localization. This temperature dependency is more pronounced when $\phi < \phi_c^p$. On the other hand, when $\phi \gg \phi_c^p$, $d\sigma/dT$ does not depend on $\phi$ and instead matches that of graphite, a fact suggesting that conduction in these samples proceeds through entangled fibers in direct contact (as supported by the nanoscale measurements in Fig. 4).

At low $\phi$, the conductivity is governed by variable range hopping (VRH) of charge carriers between localized states near the Fermi level, and takes the form[46-48]:

$$\sigma = \sigma_i + \sigma_\infty \exp\left[-\left(\frac{T_0}{T}\right)^\gamma\right], \tag{4}$$

where $\sigma_\infty$ is only weakly dependent upon temperature, and $T_0$ is a constant that depends on the electronic structure, density of states near the Fermi level, and localization length. When the density of states near the Fermi level is constant (Mott-VRH theory), the exponent $\gamma$ is related to the dimensionality $d$ of the conduction path according to[49, 50] $\gamma = (1 + d)^{-1}$. In semimetals, however, the density of states near the Fermi level may vanish and long-range Coulomb interactions between the localized electrons can open a Coulomb gap in the band structure. This so-called Coulomb-gap VRH (or Efros–Shklovskii VRH) conduction model is characterized by $\gamma = 0.5$ in all dimensions[51, 52]. Least-squares fits on 1, 2 and 2.4 vol% F2-CNF composites show that a Mott-VRH model with $\gamma = 0.25$ ($d = 3$) matches the data with a high goodness-of-fit ($R^2 > 0.997$); see Table 1.

In samples with higher $\phi$, it is expected that the transport properties are no longer controlled by the activation energy for hopping but rather by inelastic carrier-carrier and carrier-phonon interactions affecting the carriers lifetime. Neglecting the conductivity $\sigma_i$ of the matrix, such interactions yield a temperature-dependence of the form[49]

$$\sigma = \sigma_0 + AT^n, \tag{5}$$

where the value of the exponent $n$ is characteristic of the scattering interaction ($n = 0.5$ and $n > 0.75$ for carrier–carrier and carrier–phonon scattering interactions, respectively)[53]. For $\phi > \phi_c^p$, our data is well fitted by Eq. 7 (see Table 2 for the values of the exponent $n$ with $\phi$). The results suggest that the conductivity is limited by electron–phonon scattering interactions as one would expect for graphite given the low density of charge carriers.

# 7 | Conclusion

We have presented a general methodology for producing highly conductive, thermoformable oxide glass composites. Compared to previous investigations, our strategy provides a pathway to potential large-scale manufacturing of glass-CNF composites by exploiting low-cost constitutive materials and scalable fabrication processes. By optimizing the chemistry underlying the dispersion of the conductive-phase CNF in the glass, we achieve the highest room-temperature electrical conductivity to date (~ 1800 S/m), exceeding that of chemically modified glasses and surpassing that of glass composites doped with a carbon-based conductive phase. The behavior of the measured conductivity reveals clearly a transition to percolation with increased carbon-loading. Nanoscale measurements of the electrical conductivity support the existence of a connected conductive network of CNFs above the percolation threshold. The temperature dependence of the electrical conductivity of the composite samples suggests that conductivity is limited by variable range hopping between localized states at low carbon-loading and by electron-phonon scattering at high carbon-loading. Finally, Fig. 5b provides a comparison of reported values of the temperature-dependent electrical conductivities of conductive glasses.[22, 54-58] Polymer carbon composites generally provide lower conductivity and much lower operating temperatures. Our glass-CNF composites clearly exhibited higher electrical conductivity and stability over a wider range of tested temperatures. Note that the conductivity of our soda-lime-based composite is slightly lower than that of F2 lead silicate glass for the same carbon-loading fraction. We hypothesize that the higher conductivity of F2 glass is due to higher baseline conductivity of the glass matrix. Finally, we note that although the composite glass has reduced transparency due to the CNFs, they can nevertheless be formed thermally; for example, they can be thermally drawn into an electrically conductive fiber[59].

The CNF-glass composite reported here offers several accessible degrees of freedom that can be explored to engineer particular functionalities, including the volume fraction, the characteristics of the carbon fibers, the formulation of the dispersion liquid, and the dimensions and shape of the processed composite parts. These formulations present several advantages to technologies that demand small, low-weight and power-efficient (SWaP) electronic realizations. In the form of heaters, the low heat capacity of the carbon fibers improves on their response time and efficiency by facilitating heat transfer. These attributes also become essential in applications such as impulse and damping power resistors. Furthermore, the strain and temperature dependency of the conduction network can be utilized in sensing applications (strain gauge and negative temperature coefficient thermistors). At lower carbon-loading fractions, these composites may present benefits to electromagnetic shielding and electrostatic discharge applications where low-cost, oxidation resistance, and low thermal expansion are required. Our higher glass-CNF composites are candidate matrices for cathode materials in energy storage systems. For example, in rechargeable Li/S cells, S-containing compounds used as cathodes are electrically insulating thereby leading to low coulombic efficiency, and are soluble in electrolytes, which leads to capacity degradation on repeated use[60-62]. Elemental S grafted onto our porous composites with high CNF loading may help overcome such challenges while simultaneously providing better mechanical stability[60-62].

**Acknowledgments**

G. Tao acknowledges J. Qiu for encouragement, and X. Wang, R. Zheng, and Y. Wu for useful discussions. M. Molinari thanks N. B. Bercu from the LRN EA4682 for technical assistance. The authors thank Laurene Tatard for help interpreting the measurements and the personnel at the Material Characterization Facility (MCF) at the University of Central Florida for assistance. This work was supported by the MIT MRSEC through the MRSEC Program of the National Science Foundation under award number DMR-0819762, by the Australian Research Council grants CE140100003 and DP170104367. This work was performed in part at the Optofab node of the Australian National Fabrication Facility utilizing Commonwealth and South Australian State Government funding.

# Tables

**Table 1 | Least-squares fitted parameters for Eq. (4) for F2-CNF composites**

| $\phi$ (vol %) | $\sigma_i$ (S/m) | $\sigma_\infty$ (S/m) | $T_0$ (K) | $\gamma$ |
|---|---|---|---|---|
| 1 | $10^{-4}$ | $24\times10^4$ | $41\times10^6$ | 0.25 |
| 2 | $10^{-4}$ | $25\times10^4$ | $23\times10^6$ | 0.25 |
| 2.4 | $10^{-4}$ | $12\times10^4$ | $12\times10^6$ | 0.25 |

**Table 2 | Least-squares fitted parameters for Eq. (5) for F2-CNF composites**

| $\phi$ (vol %) | $\sigma_0$ (S/m) | $A$ (S/K$^n$) | $n$ |
|---|---|---|---|
| 3.7 | 2.4±0.2 | 3±2×10$^{-6}$ | 2.3±0.1 |
| 7 | 36±2 | 9±6×10$^{-5}$ | 2.1±0.1 |
| 10 | 63±9 | 1.1±2×10$^{-4}$ | 2.2±0.3 |
| 16 | 231±18 | 2.4±3×10$^{-4}$ | 2.2±0.2 |
| 36 | 955±61 | 5.3±0.05 | 1.5±0.1 |
| 46 | 1203±139 | 0.2±0.3 | 1.3±0.2 |

# Figures

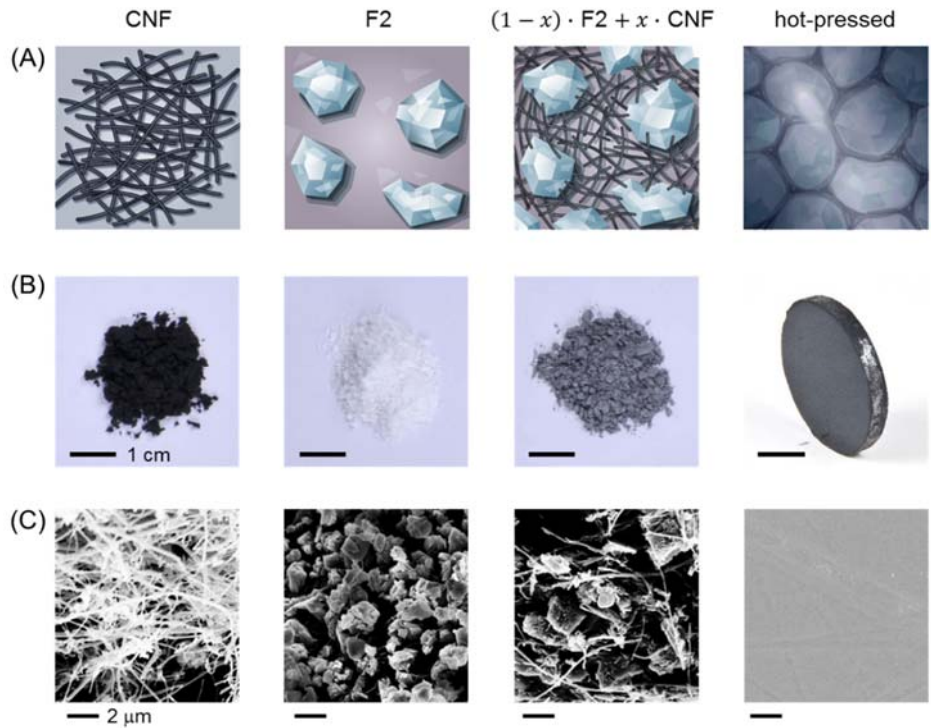

**Figure 1.** Preparation of a glass-CNF composite. (A) Schematic of the fabrication process for glass-CNF composites. From left to right: The CNF is combined with the F2 ground glass powder to prepare a mixture having the composition $(1-x) \cdot \text{F2} + x \cdot \text{CNF}$, $0 < x < 1$, which is then hot-pressed into a solid bulk. (B) Photographs of the materials; from left to right: pure CNFs, F2 glass powder, glass-CNF powder mixture after ball-milling, and pressed sample in correspondence with the steps in (A). The scale bars in the four panels are all 1 cm. (C) SEM micrographs of pure CNFs, glass powder, the glass-CNF mixture (36 vol% CNF), and surface of composite after hot press, corresponding to the panels in (B). The scale bars in the four panels are all 2 μm.

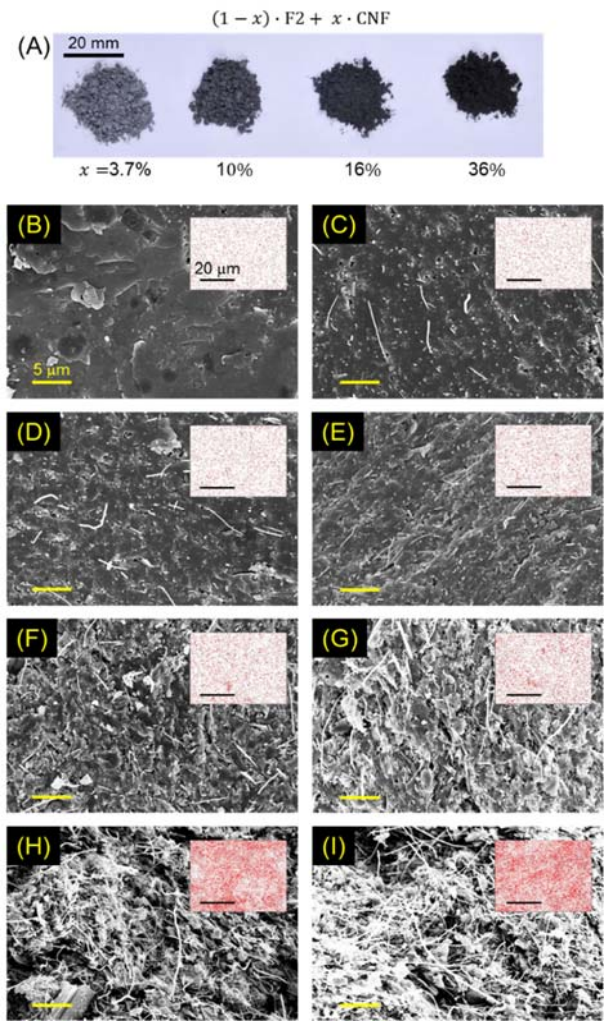

**Figure 2.** Evolution of the composite morphology with CNF loading. (A) Photographs of the glass-CNF powder mixtures with composition $(1 - x) \cdot F2 + x \cdot CNF$, where $x$ represents the carbon-loading fraction by volume $\phi$. Four examples with different values of $x$ are shown. (B)-(I) SEM micrographs of the fracture-surface microstructure for composites with CNF-loading fractions of 1, 2, 3.7, 7, 10, 16, 36, and 46 vol%. All scale bars are 5 μm. Insets show corresponding carbon-element EDS maps on the composite surface. All inset scale bars in the insets are 20 μm.

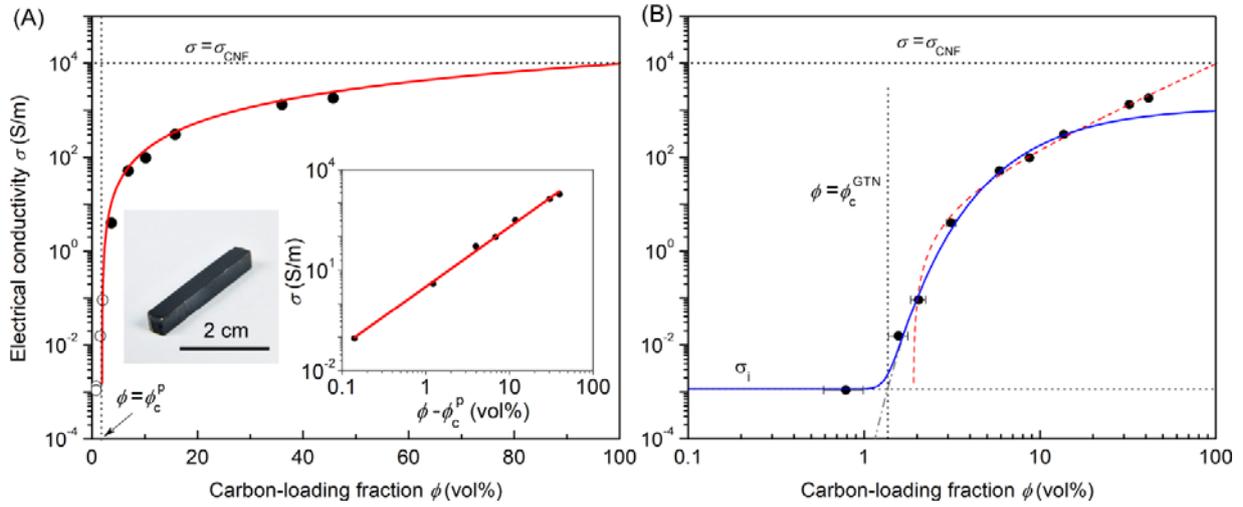

**Figure 3.** (A) Measured values of the room-temperature electrical conductivity $\sigma$ for F2-CNF composite samples with the carbon-loading fraction $\phi$. The solid red curve is a theoretical fit based on the percolation model using Eq. 1. The inset shows the good linearity of the fit on a log-log plot of $\sigma$ with $\phi - \phi_c^p$ relationship ($R^2 = 0.99$). Inset is a photograph of a typical composite sample used in the four-probe measurements. The horizontal dashed line identifies the conductivity of a CNF pellet, whereas the vertical dotted line corresponds to the percolation threshold $\phi_c^p$. (B) Fitting the data from (A) to the global tunneling network (GTN) model using Eq. 2 (solid blue curve). The curve provides an excellent fit for low $\phi$, but is less satisfactory at high $\phi$. The fit based on the percolation model (Eq. 1) is shown as a red dashed line for comparison – corresponding to the solid red curve in (A). The percolation model provides an excellent fit for high $\phi$, but is less satisfactory at low $\phi$. The horizontal dotted line represents the glass host conductivity $\sigma_i$ as determined from the GTN model, whereas the vertical dotted line corresponds to the CNF-loading fraction $\phi_c^{GTN}$ whereupon tunneling on the loose fiber network matches the matrix conductivity. We have also identified the conductivity of a CNF pellet as in (A).

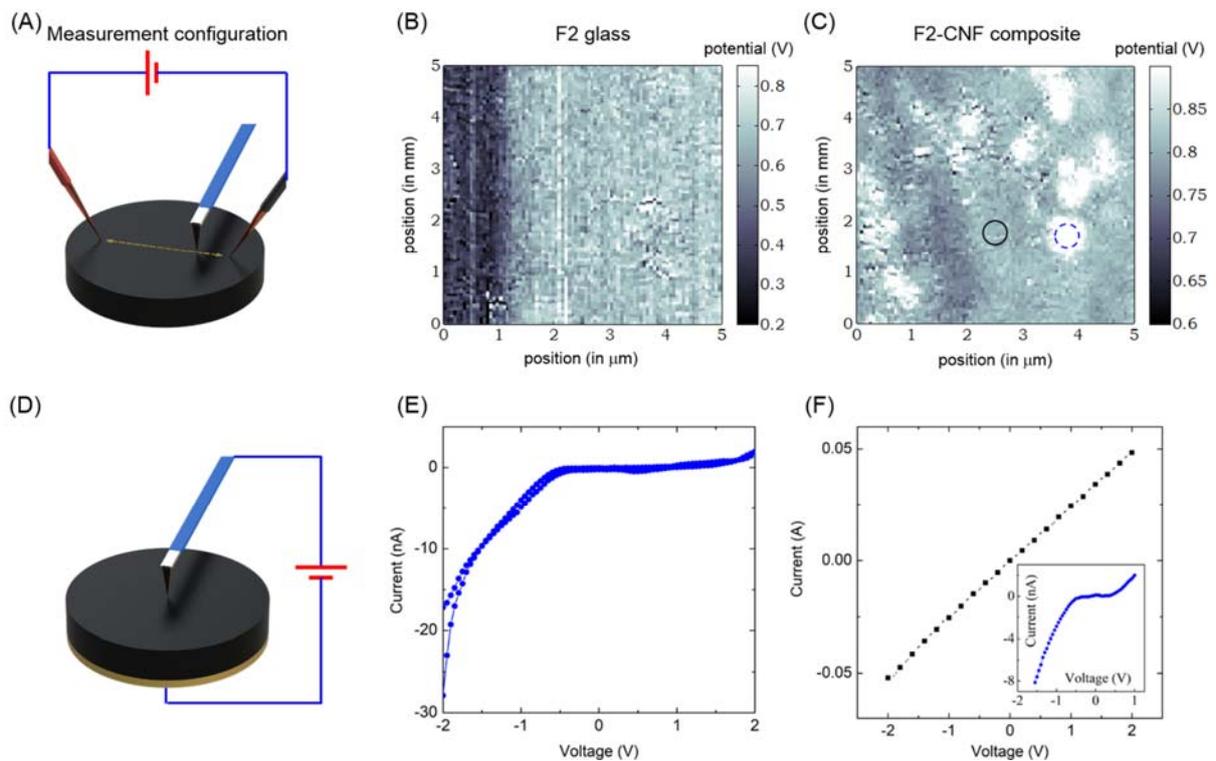

**Figure 4.** Nanoscale characterization of the electrical conductivity of the glass composite. (A) Configuration for KPFM measurements. Two contacts apply a potential across the sample and an AFM probe is scanned between them. (B) Measured potential over a 5×5 μm$^2$ area of a reference F2 glass sample that is CNF-free. A gradual drop in potential is observed along the scanning direction. (C) Measured potential over a 5×5 μm$^2$ area of a F2-CNF composite sample. The bright spots correspond to local areas of higher potential. (D) Configuration for conductive-AFM measurements. (E) I-V curves measured at two generic spots of the area shown in (B) showing non-ohmic behavior. (F) I-V curve measured at a bright spot (the dashed blue circle) in (C) revealing ohmic behavior. Inset is the I-V curve measured at a spot in the dielectric (black circle) in (C). AFM tip diameter is 40 nm.

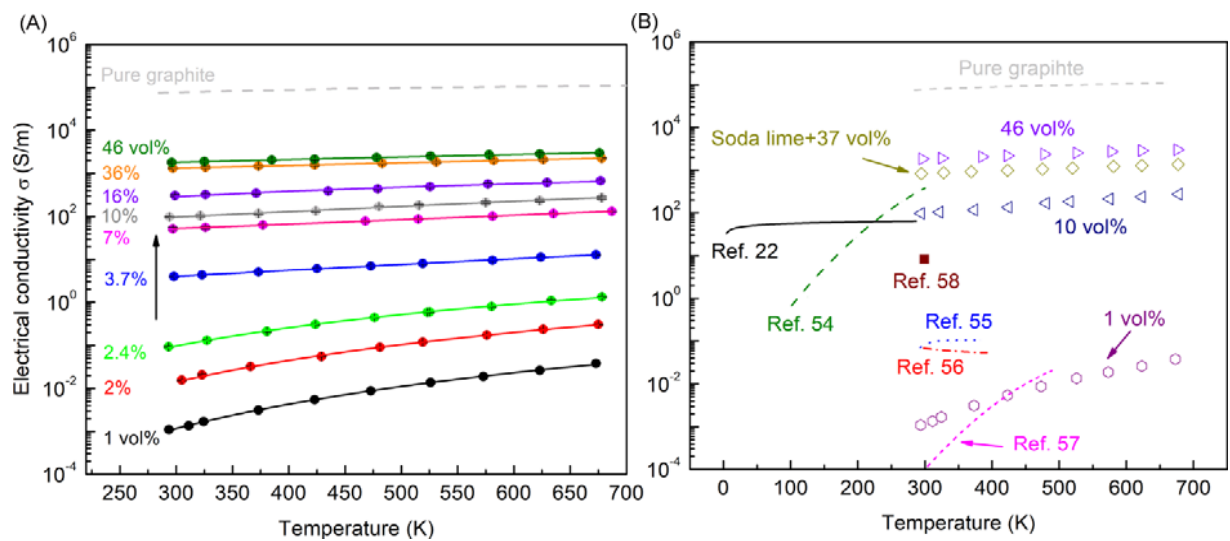

**Figure 5.** (A) Measured electrical conductivity of F2-CNF composites as a function of temperature at various carbon-loading fractions (1 vol% to 46 vol% from bottom to top). The conductivity of samples with $\phi < \phi_c^p$ is fitted with the Mott-VRH model (Eq. 4), while that of samples with $\phi > \phi_c^p$ is fitted with Eq. 5. The temperature-dependent conductivity of bulk polycrystalline graphite is represented by the grey dashed line. (B) Comparison of the electrically conductivity of the F2-CNF composite developed here with previous reports of conductive glassy materials. The results for the F2+CNF composite are shown as symbols (triangles or circles), corresponding to CNF-loading fractions of 1 vol%, 10 vol%, and 46 vol% selected from (A). We have also plotted the measured values for a soda-lime glass-CNF composite (see Supplemental Material). The other data are as follows: $SiO_2$+10 vol% CNT[22], $Cu27.5Ge2.5Te70$[54], Epoxy+2.5 wt% CNT[55], Epoxy+8 wt% Graphite[56], Phosphate+50 mol% $V_2O_3$[57], Borosilicate+10 wt%CNT[58].